# $Bi_2Te_3$/Si thermophotovoltaic cells converting low temperature radiation into electricity

Xiaojian Li[1], Chaogang Lou[1][1], Xin Li[1], Yujie Zhang[1], Zongkai Liu[1], Bo Yin[2]

[1]*Joint International Research Laboratory of Information Display and Visualization, School of Electronic Science and Engineering, Southeast University, Nanjing, 210096, P. R. China*

[2]*School of Physics and Electronics, Nanyang Normal University, Nanyang, 473007, P. R. China*

The thermophotovoltaic cells which convert the low temperature radiation into electricity are of significance due to their potential applications in many fields. In this work, $Bi_2Te_3$/Si thermophotovoltaic cells which work under the radiation from the blackbody with the temperature of 300 K-480 K are presented. The experimental results show that the cells can output electricity even under the radiation temperature of 300 K. The band structure of $Bi_2Te_3$/Si heterojunctions and the defects in $Bi_2Te_3$ thin films lower the conversion efficiency of the cells. It is also demonstrated that the resistivity of Si and the thickness of $Bi_2Te_3$ thin films have important effects on $Bi_2Te_3$/Si thermophotovoltaic cells. Although the cells' output power is small, this work provides a possible way to utilize the low temperature radiation.

Thermophotovoltaic cells have attracted much attention in recent years because of their ability to convert thermal energy into electrical energy directly [1-8]. Different from the solar cells which can only work under the illumination of visible and near-infrared light, the thermophotovoltaic cells can convert long wavelength infrared radiation into electricity. Owing to this feature, this kind of cells are able to be applied extensively. Therefore, a lot of efforts have been paid in this area and many good progresses have been made [9-15].

So far, most of the researches about the thermophotovoltaic cells focus on those which work well under the high temperature radiation [16-19]. There are few reports about the cells which can absorb the infrared photons radiated from low temperature (below 500 K) heat sources. In fact, because the low temperature radiation exists more widely than the high temperature one, the thermophotovoltaic cells working under the low temperature radiation have more extensive applications.

To make use of the low temperature radiation, in our previous work [20], the thermophotovoltaic cells consisting of n-type $Bi_2Te_3$ thin films and p-type $Sb_2Te_3$ thin films were proposed. This kind of cells can absorb the infrared photons radiated from the heat sources with the temperature 300-470 K. However, the small difference between the Fermi levels of $Bi_2Te_3$ and $Sb_2Te_3$ thin films makes the p-n junction similar to a resistor and the cells' output power is very small.

To improve the performance of the cells, in this work we demonstrate the thermophotovoltaic cells which include $Bi_2Te_3$/Si heterojunctions by substituting the p-type $Sb_2Te_3$ thin films with p-type single crystalline Si wafers. Because the difference between the Fermi levels of $Bi_2Te_3$ thin films and Si is about 0.67 eV-0.89 eV (much higher than that between $Bi_2Te_3$ and $Sb_2Te_3$ thin films), the introduction of Si brings the cells a built-in electric field which is strong enough to separate the photogenerated electron-hole pairs. Here, the output current of $Bi_2Te_3$/Si thermophotovoltaic cells under the radiation from a low-temperature heat source (300-480 K) is explored. The effects of Si wafer's doping concentration and the thickness of $Bi_2Te_3$ thin films on $Bi_2Te_3$/Si thermophotovoltaic cells are also discussed. This kind of cells provide a convenient method to convert low temperature radiation into electricity and to broaden the application of thermophotovoltaic cells.

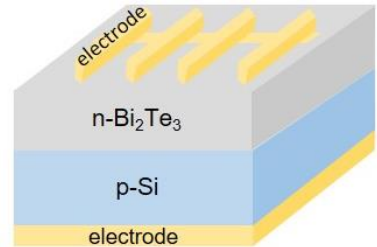

FIG. 1. The schematic structure of $Bi_2Te_3$/Si thermophotovoltaic cells.

The schematic structure of $Bi_2Te_3$/Si thermophotovoltaic cells is shown in Fig. 1. In the experiments, 100 nm Ag films were at first deposited on the back face of pre-cleaned p-type Si substrates (2 cm×2 cm) as the bottom electrode by vacuum thermal evaporation. Then, $Bi_2Te_3$ thin films were deposited on the front face of Si substrates by RF magnetron sputtering. RF power for depositing $Bi_2Te_3$ thin films was 100 W, and the sputtering time was set as 300 s. Finally, 150 nm Ag films as the top electrode were deposited on $Bi_2Te_3$ thin films by vacuum thermal evaporation.

Figure 2(a) shows the SEM image of top view of $Bi_2Te_3$ thin films which are polycrystalline and consist of the grains with the size from dozens of nanometers to one hundred nanometers. The atomic ratio of $Bi_2Te_3$ thin films is Bi:Te=39.35:60.65, which agrees with the stoichiometric composition. Figure 2(b) shows the cross-section of $Bi_2Te_3$/Si heterostructures. The thickness of $Bi_2Te_3$ thin films is about 310 nm.

---

[1] lcg@seu.edu.cn

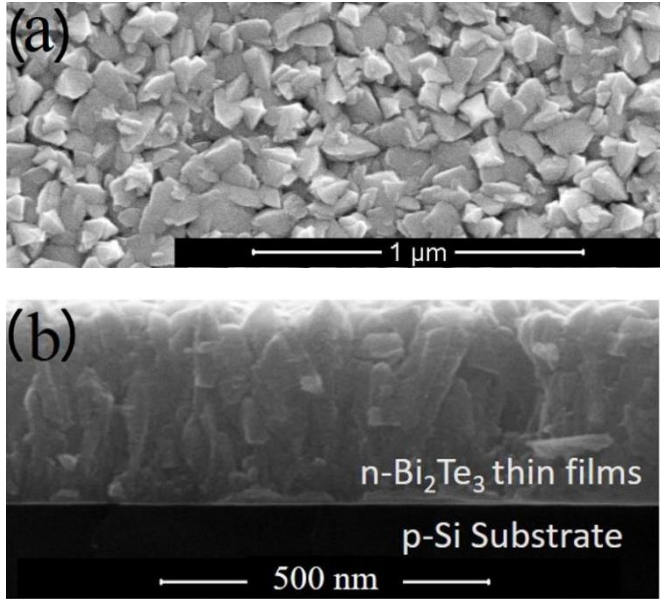

FIG. 2 SEM images of (a) the surface morphology of $Bi_2Te_3$ thin films and (b) the cross-section of $Bi_2Te_3$/Si heterojunctions.

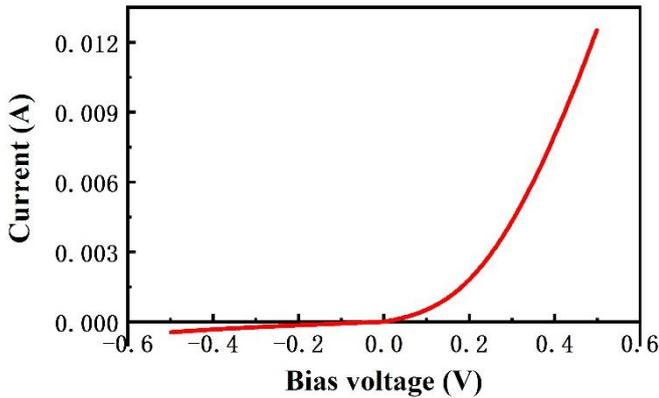

FIG. 3 I-V curve of $Bi_2Te_3$/Si (the resistivity of Si is 0.001 Ωcm) heterojunctions at room temperature.

Figure 3 shows the I-V curve of a $Bi_2Te_3$/Si (the resistivity of Si is 0.001 Ωcm) heterojunction at room temperature. Different from the previous $Bi_2Te_3$/$Sb_2Te_3$ heterojunction whose I-V curve looks like a straight line [20], the I-V curve of $Bi_2Te_3$/Si heterojunction verifies the formation of the p-n heterojunction.

In general, the measurement of the external quantum efficiency is necessary for a thermophotovoltaic cell so that its response to the infrared photons with different wavelength can be known. However, for $Bi_2Te_3$/Si thermophotovoltaic cells, the lack of control detectors working in the long wavelength range makes the test difficult. Here, to know the spectral response of the cells, the output current under different wavelength radiation is tested instead of measuring the external quantum efficiency, as shown in Fig. 4(a). The cells can generate photocurrent by absorbing the infrared radiation with the wavelength shorter than 8.2 μm, which agrees with the bandgap of $Bi_2Te_3$ (0.15-0.17 eV) reported in literatures [21-23].

The reflection and transmission of the cells in the wavelength range of 2.5-8 μm are given in Fig. 4(b). It can be calculated that the average reflection over the whole wavelength range is around 46%. In the range from 2.5 to 4.2 μm, the transmission increases with the wavelength while the absorption decreases from 42.8% to 22.6%. In the range longer than 4.2 μm, the transmission decreases slightly with the wavelength and the absorption varies slightly with the average value around 24%.

The current density-voltage (J-V) characteristics for $Bi_2Te_3$/Si thermophotovoltaic cells under the radiation from the blackbody (with an emissivity of 0.95) with different temperature are shown in Fig. 4(c). Both $V_{OC}$ and $I_{SC}$ increase with the temperature of the radiation source. When the temperature rises from 300 K to 480 K, $V_{OC}$ increases from 0.095 mV to 1.756 mV and $I_{SC}$ increases from 7.383 μA to 104.693 μA.

The variation of the cells' output can be attributed to the increase in the number of absorbable photons. According to Planck's Law [24, 25], as the temperature of the radiation source increases, the peak wavelength of the radiation shifts from 9.66 μm at 300 K to 6.04 μm at 480 K. This makes both the fraction of the absorbable photons and the total number of the absorbable infrared photons rise. So, more carriers are excited and collected, and make $V_{OC}$ and $I_{SC}$ increase.

The fill factor of the cells can be calculated from Fig. 4(c). It hardly varies with the radiation source temperature and is about 25.6%. This value is much lower than that of single crystalline Si solar cells, which is currently around 80%. By including the radiation data of the blackbody source (from Planck's Law) and the absorption data of $Bi_2Te_3$ thin films (from Fig. 4(b)), the conversion efficiency (η) of the cells under different radiation temperatures can be calculated approximately, as shown in Fig. 4(d). Clearly, the cell's efficiency increases with the radiation source temperature. This might be attributed to the increasing fraction of the absorbable photons and their higher absorption by $Bi_2Te_3$ thin films.

Compared with $Bi_2Te_3$/$Sb_2Te_3$ cells [20], $Bi_2Te_3$/Si cells have much better performance: the efficiency rises from ~

$10^{-8}$ to ~ $10^{-6}$. The improvement results from the bigger difference between the Fermi levels of $Bi_2Te_3$ and Si, which makes the excited electron-hole pairs separated more rapidly and collected more efficiently.

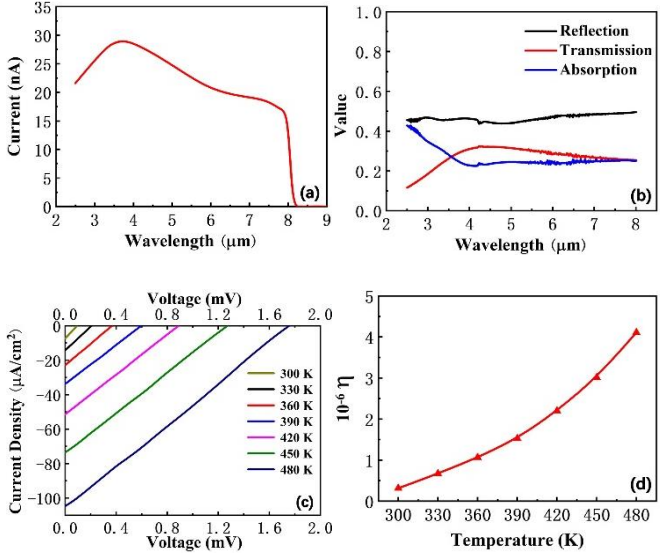

FIG. 4 (a) Relationship between the output current of the $Bi_2Te_3$/Si thermophotovoltaic cells and the wavelength (measured in the dark environment). (b) The reflection, transmission and absorption of $Bi_2Te_3$/Si thermophotovoltaic cells. (c) Current density-voltage (J-V) characteristics. (d) Calculated efficiency ($\eta$) for $Bi_2Te_3$/Si thermophotovoltaic cells under the radiation source of the temperature from 300 K to 480 K.

However, the efficiency of $Bi_2Te_3$/Si cells is still too low, less than $5\times10^{-6}$. There are two possible reasons which are mainly responsible for this: one is the potential barrier at the interface of the heterojunction and the other is the carrier recombination inside $Bi_2Te_3$ thin films and on their surface.

Figure 5(a) shows the spectrum from the ultraviolet photoelectron spectrometer, from which the work function of $Bi_2Te_3$ thin films can be calculated [26]. The value of the work function is 4.35 eV, which is equal to the difference between incident photon's energy (21.2 eV) and the secondary-electron cut-off energy (16.85 eV). Because the bandgap of n-type $Bi_2Te_3$ is narrow, the value of the electron affinity of $Bi_2Te_3$ should be very close to the work function. From the data of the electron affinity of $Bi_2Te_3$ (around 4.35 eV), the band gap of $Bi_2Te_3$ (0.15 eV), the electron affinity of Si (4.05 eV) and the bandgap of Si (1.12 eV), the schematic band diagram of $Bi_2Te_3$/Si heterojunction can be plotted, as shown in Fig. 5(b).

Because the infrared photons from the radiation source cannot be absorbed by Si, all of the electron-hole pairs are generated only in $Bi_2Te_3$ thin films. It can be seen in Fig. 5(b) that there is not any barrier for photo-generated electrons to move to collector (on the left side), while, for photo-generated holes in $Bi_2Te_3$ thin films, the existence of the valence band offset at the interface prevents them from entering into p-type region. The holes can pass through the barrier only by tunneling. This reduces the output current of $Bi_2Te_3$/Si thermophotovoltaic cells. The possible solution for this problem is to find the suitable materials whose energy levels match with those of $Bi_2Te_3$ so that the barrier of the holes can be reduced.

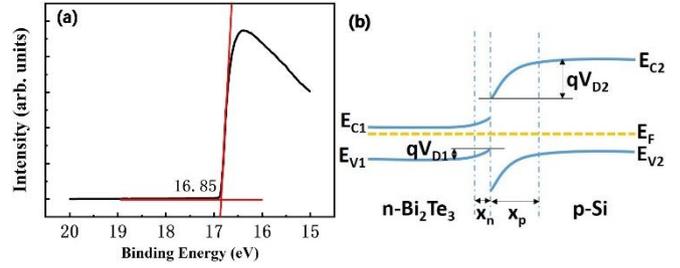

FIG. 5 (a) Ultraviolet photoelectron spectrometer's spectrum of $Bi_2Te_3$ thin films. (b) The band diagram for the $Bi_2Te_3$/Si heterojunction.

Another possible reason responsible for the low efficiency is the defects in $Bi_2Te_3$ thin films. The grain boundaries and antisite defects of the polycrystalline thin films leads to serious carrier recombination which reduces the output power [27, 28]. Therefore, improving the quality of the thin films and passivating their surfaces are important for the enhancement of the efficiency of the cells.

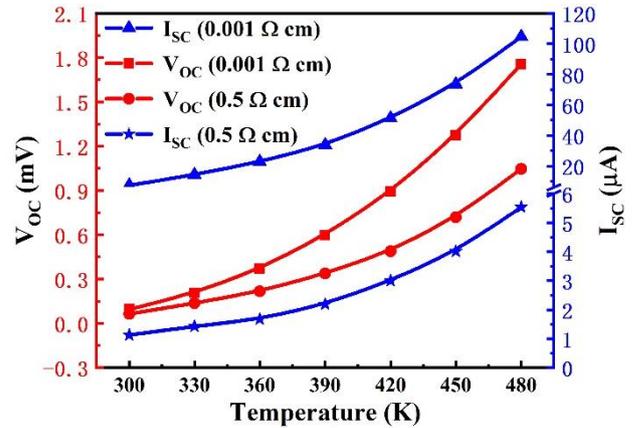

FIG. 6 $V_{OC}$ and $I_{SC}$ of $Bi_2Te_3$/Si thermophotovoltaic cells with Si of different resistivity (0.5 $\Omega$cm and 0.001 $\Omega$cm).

To understand better $Bi_2Te_3$/Si thermophotovoltaic cells, the effects of resistivity of Si on the cells is investigated.

Figure 6 shows $V_{OC}$ and $I_{SC}$ of the cells with Si of the resistivity 0.5 Ωcm and 0.001 Ωcm. It can be seen that both $V_{OC}$ and $I_{SC}$ of the cells with Si of 0.5 Ωcm are lower than those of the cells with Si of 0.001 Ωcm.

This might be attributed to the different width of the heterojunction's depletion region, which depends on the doping concentration of Si. The n-$Bi_2Te_3$ thin films are self-doped by vacancy and antisite defects [29], and the doping concentration is around $10^{20}$ cm$^{-3}$ (measured on HMS-5000) [20]. The doping concentrations of Si are $3\times10^{16}$ cm$^{-3}$ (0.5 Ωcm) and $1.5\times10^{20}$ cm$^{-3}$ (0.001 Ωcm), respectively. When the heterojunction is formed, the width $x_n$ (shown in Fig. 5(b)) of the depletion region in $Bi_2Te_3$ thin films depends on the difference between the doping concentrations of $Bi_2Te_3$ and Si [30, 31]. This makes $x_n$ of the cells with Si of 0.001 Ωcm larger than that with Si of 0.5 Ωcm, so more infrared photons can be absorbed in the depletion region of the cells with Si of 0.001 Ωcm.

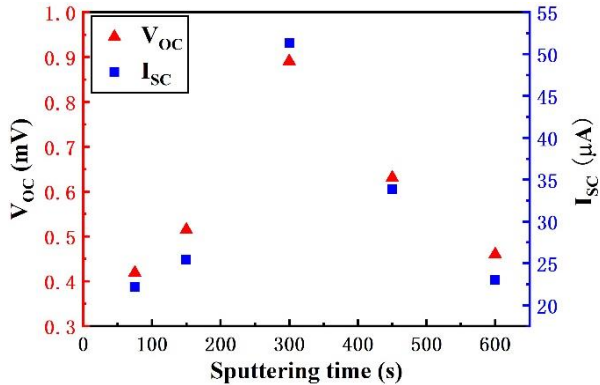

FIG. 7 $V_{OC}$ and $I_{SC}$ of $Bi_2Te_3$/Si (0.001 Ωcm) thermophotovoltaic cells with different sputtering times of $Bi_2Te_3$ thin films under the radiation from the heat source at 420 K.

The effects of thickness of $Bi_2Te_3$ thin films on the thermophotovoltaic cells is also investigated. Figure 7 shows $V_{OC}$ and $I_{SC}$ of $Bi_2Te_3$/Si (the resistivity of Si is 0.001 Ω cm) thermophotovoltaic cells under the radiation of the source at 420 K. In these cells, $Bi_2Te_3$ thin films have different thickness and their sputtering time is set as 75 s, 150 s, 300 s, 450 s, and 600 s, respectively. As the sputtering time increases, that is, the thickness of $Bi_2Te_3$ thin films increases, $V_{OC}$ and $I_{SC}$ of the cells keep rising until 300 s. After that, both of them turn to decrease. The maximum values appear at 300 s, which corresponds to the thin film thickness of about 310 nm.

On the basis of the data in Fig. 7, it is natural to think that the optimal thickness 310 nm might be the width of the depletion region in $Bi_2Te_3$ thin films. This is because, if the thickness of $Bi_2Te_3$ thin films is below 310 nm, the cells absorb less radiation and the output power becomes smaller. While, if the thickness of $Bi_2Te_3$ thin films is larger than 310 nm, the excited electrons have to travel over a longer distance and have more chances to be recombined. However, according to classical theory [32], the calculated width of the depletion region in $Bi_2Te_3$ thin film is only 1.9 nm, much smaller than 310 nm. It might be attributed to the large number of intrinsic carriers and defects which make the transport and recombination of carriers in $Bi_2Te_3$ thin films complicated and the calculation of width of depletion region no longer obey the classical theory. No matter what the reason for the disagreement is, this topic is interesting and worth further investigation.

Although the conversion efficiency of $Bi_2Te_3$/Si thermophotovoltaic cells is low and the output power is small, the results show that the low temperature radiation can be converted into electricity. Especially, when the temperature of the radiation source is 300 K, the cells can still convert the weak radiation into electricity. This is the advantage of $Bi_2Te_3$/Si thermophotovoltaic cells which provides a convenient method to make use of low-temperature radiation and demonstrates that obtaining energy from room-temperature surroundings is possible.

Furthermore, compared with currently used thermophotovoltaic materials like GaSb, InSb, InAs, etc., $Bi_2Te_3$ is more inexpensive and its thin films are easy to be fabricated by vacuum deposition. In addition, reducing the distance between the cells and the heat source can improve their performance further due to near-field effect [9, 33]. These give $Bi_2Te_3$/Si thermophotovoltaic cells more opportunities to be applied in many fields.

As mentioned above, $Bi_2Te_3$/Si thermophotovoltaic cells have a low conversion efficiency which results from the valence band offset at the interface and the carrier recombination in the polycrystalline $Bi_2Te_3$ thin films. Therefore, substituting Si with the semiconducting materials of a suitable energy band might lower the potential barrier at the interface and improve the transport of holes. To reduce the carrier recombination, it is necessary to decrease the number of defects by improving the crystallinity of $Bi_2Te_3$ thin films and passivating their surface.

Another challenge which $Bi_2Te_3$/Si thermophotovoltaic cells face is how to enhance the absorption of the long wavelength photons by $Bi_2Te_3$ thin films. Fig. 4(b) shows that, for the $Bi_2Te_3$ thin films of 310 nm thickness, most of the radiated energy is reflected or transmitted. In order to improve the absorption, the possible way is to increase the thickness of the thin films. However, this will raise the carrier recombination and reduce the output voltage, as shown in Fig. 7.

These problems restrict seriously the performance of $Bi_2Te_3$/Si thermophotovoltaic cells. But, on the other hand, they also indicate that a big potential improvement is possible. Therefore, our next efforts will focus on finding the solutions to enhance the cells.

In conclusion, it has been demonstrated that $Bi_2Te_3$/Si thermophotovoltaic cells can output electrical power by absorbing the radiation from the heat source with temperature from 300 K to 480 K. The barrier at the interface of the junction, the defects in $Bi_2Te_3$ thin films and the less absorption reduce seriously the output power. The higher doping concentration of Si leads to a higher output of the cells, and the optimal thickness of $Bi_2Te_3$ thin films is around 310 nm when the resistivity of Si is 0.001 Ωcm, which might also be the width of the depletion region in $Bi_2Te_3$ thin films. Although there exist still some challenges, this work provides a possible method to generate electricity through absorbing the radiation from low (or room) temperature surroundings.

## ACKNOWLEDGMENTS

The authors thank the supports from the Natural Science Foundation of Jiangsu (Grant No. BK2011033) and the Primary Research & Development Plan of Jiangsu Province (Grant No. BE2016175).